\documentclass[reprint,aps,pra,superscriptaddress]{revtex4-1}

\usepackage{amsmath,amsthm,amssymb}
\usepackage{mathtools}
\usepackage{units}

\usepackage{url}
\usepackage{comment}
\usepackage{dsfont} 

\usepackage{graphicx}
\usepackage{bm}
\usepackage{color}
\usepackage{braket}
\usepackage{hyperref}
\usepackage[utf8]{inputenc}
\hypersetup{colorlinks=true,linkcolor=blue,citecolor=blue}

\newcommand{\h}{\mathcal{H}}
\newcommand{\he}{\mathcal{H}^{(\mathrm{eff})}}

\begin{document}
	
	\title{Topological edge-states of the $\mathcal{PT}$-symmetric Su-Schrieffer-Heeger model:\\
	An effective two-state description}	
	
	\author{A. F. Tzortzakakis}
	\affiliation{Institute of Electronic Structure and Laser, FORTH, GR-70013 Heraklion, Crete, Greece}
	\affiliation{Department of Physics, National and Kapodistrian University of Athens, GR-15784 Athens, Greece}
	
	\author{A. Katsaris}
	\affiliation{Department of Physics, National and Kapodistrian University of Athens, GR-15784 Athens, Greece}
	
	\author{N. E. Palaiodimopoulos}
	\affiliation{Institute of Electronic Structure and Laser, FORTH, GR-70013 Heraklion, Crete, Greece}
	
	\author{P.~A.~Kalozoumis}
	\affiliation{Department of Engineering and Informatics, Hellenic American University, 436 Amherst Street, Nashua, NH 03063, USA}
	
	\author{G. Theocharis}
	\affiliation{LAUM, CNRS, Le Mans Université, Avenue Olivier Messiaen, 72085 Le Mans, France}
	
	\author{F. K. Diakonos}
	\affiliation{Department of Physics, National and Kapodistrian University of Athens, GR-15784 Athens, Greece}
	
	\author{D. Petrosyan}
	\affiliation{Institute of Electronic Structure and Laser, FORTH, GR-70013 Heraklion, Crete, Greece}
	\affiliation{A.  Alikhanyan  National  Science  Laboratory  (YerPhI),  0036  Yerevan,  Armenia}
	
	\begin{abstract}
	We consider the non-Hermitian, parity-time $(\mathcal{PT})$ symmetric extensions of the one-dimensional Su-Schrieffer-Heeger (SSH) model in the topological non-trivial configuration. We study the properties of the topologically protected edge states, and  develop an effective two-state analytical description of the system that accurately predicts the $\mathcal{PT}$-symmetry breaking point for the edge states. 
	We verify our analytical results by exact numerical calculations.
	\end{abstract}
	
	\date{\today}
	
	\maketitle
	
	\section{Introduction}
	
	Following the discovery of the quantum Hall effect, the notion of topological order was used extensively as a new criterion for classifying distinct quantum phases of matter \cite{QHE1,QHE2,QHE3}. One of the most significant achievements in this field was the prediction and observation of topological insulators -- electronic materials that support conducting states localized on their surface/edges despite being insulating in their interior due to the existence of a bulk energy gap \cite{Review1,Review2}. This property is a consequence of the combined effect of topological order and symmetry protection which provides extraordinary robustness to disorder and external perturbations. Topological insulators are of fundamental significance and they could lead to various potential applications, such as high-performance electronic and robust spintronic devices, protection from decoherence in quantum computing, and many others \cite{Appl0,Appl1,Appl2,Appl3}.
	
	Recently, a classical counterpart of topological insulators was proposed \cite{EM1,EM2} and realized \cite{EM3} using photonic devises. The classical nature of optical platforms makes experimental observation and manipulation of electromagnetic waves substantially easier, preserving at the same time most of the engaging physics related to topology and symmetry protection. This extension then offers new perspectives in the broad field of topological physics \cite{EMRev,EMRev2,EMRev3}. 
	
	In the meantime, another symmetry based phenomenon -- that of parity-time ($\mathcal{PT}$) symmetry and exceptional points -- has 
	attracted much attention in optical physics \cite{PT1,PT2,PT3,PT4,PT5}. Specifically, it was recently realized that optical amplification (gain) and dissipation (loss) can be employed to implement complex potentials and explore the counterintuitive physics of non-Hermitian effects. If, in addition, this optical gain and loss is incorporated in a balanced, antisymmetric manner that respects $\mathcal{PT}$-symmetry, the corresponding system can possess completely real spectrum despite being non-Hermitian \cite{PT0}. Then, by fine-tuning the gain-loss contrast, one can actualize spontaneous $\mathcal{PT}$-symmetry breaking in the system, whereby its eigenvalues coalesce into exceptional points (non-Hermitian degeneracies) and turn from real to complex. 
	
	It is then natural to combine the two effects and explore their interplay. Due to their seemingly contradictory character, with topology inherently related to robustness and non-Hermiticity and exceptional points to extreme sensitivity, it was initially debated whether the two phenomena could coexist \cite{NhTopoQ,Esaki,Schomerus}. But recent theoretical and experimental studies have shown that the combined effect of topology and non-Hermiticity yields an even more exotic and unexpected physical behavior \cite{NhTopo1,NhTopo2,Makris1,NhTopo25,NhTopo3,Breakup,NhTopo4,Makris2,NhTopo5,NhTopo6}, featuring non-Hermitian topological light funneling \cite{NhTopoFun} and steering \cite{NhTopoSteer}, as well as topological lasing \cite{TopoLaser1,TopoLaser2,TopoLaser3}.
	
	In this paper, we focus on the edge-state properties of the $\mathcal{PT}$-symmetric Su-Schrieffer–Heeger (SSH)
	model, which is the simplest system where $\mathcal{PT}$-symmetry and topology can coexist. Using an appropriate ansatz for the edge-state wavefunctions, we derive an effective two-state model Hamiltonian that includes the coupling between the edge states and their (complex) energies. This model then leads to approximate analytical expressions for the values of the gain-loss contrast at the $\mathcal{PT}$-symmetry breaking (exceptional) point of the edge-state eigenvalues at zero energy. This so-called recovery of zero-modes by non-Hermitian means can strengthen the topologically protected characteristics of the edge states and was recently demonstrated experimentally in finite SSH optical lattices \cite{Breakup}. 
	We finally show that, under appropriate conditions, our effective description can accurately predict the exceptional points for a large variety of non-Hermitian potentials as long as they respect the basic symmetries of the model and retain its bulk gap open. In all cases, we verify our analytical results with exact numerical calculations and discuss the extent of our approach.
	
	The paper is organized as follows.
	In Sec.~\ref{Sec:Hermitian} we review the properties of the Hermitian SSH model focusing on its topological edge states. 
	In Sec~\ref{Sec:PT} we study the $\mathcal{PT}$-symmetric SSH model with uniform gain-loss contrast, derive the effective edge-state Hamiltonian and compare its analytical predictions for the exceptional-point position and edge-state wavefunctions with exact numerical calculations. In Sec.~\ref{Sec:Random} we generalize our approach to a SSH model with spatially-inhomogeneous but globally $\mathcal{PT}$-symmetric gain-loss contrast. 
	Our conclusions are summarized in Sec.~\ref{Sec:Concl}.

	\section{Hermitian SSH model}
	\label{Sec:Hermitian}
	
\begin{figure}[tb]
		\centering
		\includegraphics[clip,width=\columnwidth]{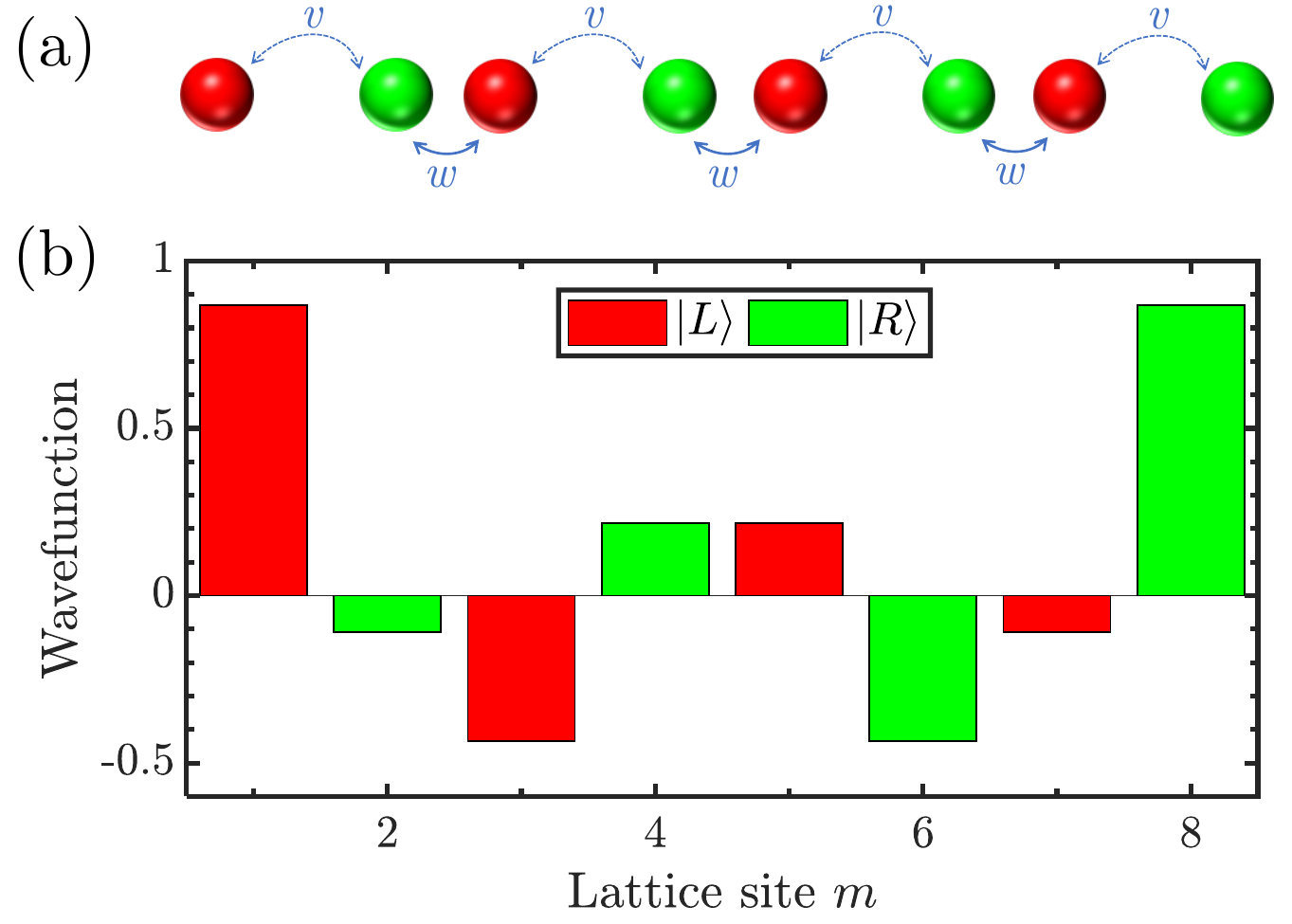}
		\caption{(a) Schematic of an SSH lattice of $M=8$ sites ($N=4$ unit cells) with alternating nearest-neighbor couplings $v<w$ (topologically non-trivial configuration), as described by Hamiltonian (\ref{ham}).
		(b) Wavefunction profiles of the ansatz edge states $\ket{L,R}$ of Eqs.~(\ref{ans})  for $u=w/v=2$.}
		\label{Fig.1}
	\end{figure}

	We consider a one-dimensional lattice of $M=2N$ sites with staggered hopping amplitudes $v,w\ge 0$ between the nearest neighbors. The system thus consists of two sublattices forming $N$ unit cells, as shown in Fig.~\ref{Fig.1}(a). 
	The ``single-particle" Hamiltonian of the system is 
	\begin{equation}
		\begin{split}			 
			\h&=v\sum_{m_\mathrm{odd}=1}^{M-1}\big[\ket{m}\bra{m+1}+\mathrm{h.c.}\big]\\
			&+w\sum_{m_\mathrm{even}=2}^{M-2}\big[\ket{m}\bra{m+1}+\mathrm{h.c.}\big].
			\label{ham}
		\end{split}
	\end{equation}
	
	Since sites of each sublattice are coupled only with sites of the other sublattice, the system possesses chiral (sublattice) symmetry. This symmetry is formally represented by the operator $\Sigma_z\equiv \mathds{1}_N\otimes \sigma_z$, where $\mathds{1}_N$ is the $N\times N$ identity operator and $\sigma_z$ is the Pauli matrix for a unit cell. Due to the chiral symmetry, this operator anticommutes with the Hamiltonian, $\{\Sigma_z,\h\}=0$, whose spectrum is symmetric around zero, $E\to-E$. 
	
	\subsection{Bulk states}
	
	The bulk $k$-space Hamiltonian of the system is
	\begin{equation}
		\h(k)=\begin{pmatrix}
			0 & v+w e^{-ik} \\
			v+w e^{ik} & 0
		\end{pmatrix}\equiv \begin{pmatrix}
			0 & h(k) \\
			h^*(k) & 0
		\end{pmatrix},
	\end{equation}
	which yields the dispersion relation
	\begin{equation}
		E(k)=\pm|h(k)|=\pm\sqrt{v^2+w^2+2vw\cos{k}}
	\end{equation}
	for a pair of bands with an energy gap $E_g=2|v-w|$. It has been shown that the topological invariant of the SSH model is a winding number \cite{Esaki}
	\begin{equation}
		\mathcal{W} =\dfrac{1}{2\pi i} \int_{-\pi}^{\pi} dk  \dfrac{d}{dk}\log{h(k)} =\dfrac{1}{2\pi i} \oint_{|z|=1} \dfrac{dz}{v/w+z}.
		\label{eq:Windnum}
	\end{equation}
	It then follows from the Cauchy's integral theorem  that the ratio  
	$u \equiv w/v$ determines the topological properties of the system: 
	for $u>1$ the winding number is nonzero ($\mathcal{W}=1$) and the system is topologically nontrivial, possessing zero-energy, topologically protected edge states, while for $u<1$ the system is topologically trivial ($\mathcal{W}=0$) and all the states are the bulk states in the two bands with the positive and negative energies. At $u=1$ ($v=w$) the gap vanishes and the spectrum reduces to that of a uniform lattice, $E(k) = -2w \cos{k/2}$ ($k \in [-2\pi,2\pi]$).  
	
	\subsection{Edge states}
	
	We now assume the topologically nontrivial regime $u>1$ and consider the two edge states. Following the analysis in Refs.~\cite{Book,Efrem,Wang,Lang2018}, we present an effective description of the edge states that will be used in the following sections. In the thermodynamic limit, the left $\ket{L}$ and right $\ket{R}$ edge states are zero-energy eigenstates of the Hamiltonian,
	\begin{equation}
		\h \ket{L,R}=0 \quad (M=2N \to \infty). \label{eq:ham0}
	\end{equation} 
	Substituting here the Hamiltonian (\ref{ham}), we obtain simple recurrence relations for the amplitudes $c_m^{L,R}$ of the wavefunctions $\ket{L,R} = \sum_{m} c_m^{L,R} \ket{m}$ as
	\begin{subequations}
		\begin{eqnarray}
			c_{m+2}^{L,R} = -c_{m}^{L,R}/u  \quad (m \, \mathrm{odd}), \\
			c_m^{R,L} =- c_{m+2}^{R,L}/u \quad (m \, \mathrm{even}) .
		\end{eqnarray}
		\label{tb}
	\end{subequations}
	Assuming that $\ket{L}$ has support only on the odd sublattice sites ($c_m^L=0 \, \forall \, m \, \mathrm{even}$), and $\ket{R}$ has support only on the even sublattice sites ($c_m^R=0 \,  \forall \, m \, \mathrm{odd}$), the relations (\ref{tb}) result in the  wavefunctions
	\begin{subequations}
		\begin{eqnarray}
			\ket{L} &=& c_L \sum_{m_{\mathrm{odd}}=1}^{M-1}(-u)^{-(m-1)/2}\ket{m} , 
			\\
			\ket{R} &=& c_R \sum_{m_{\mathrm{even}}=2}^{M}(-u)^{-(M-m)/2}\ket{m} ,
		\end{eqnarray}
		\label{ans}
	\end{subequations}
	where $c_L = c_1^{L}$ and $c_R = c_M^{R}$ are given by 
	\begin{equation}
		|c_L|^2=|c_R|^2=\dfrac{1-u^{-2}}{1-u^{-M}},
		\label{eq:clr}
	\end{equation}
	and, for simplicity, can be assumed real. 
	Even though for any finite system the states in Eq.~(\ref{ans}) are not exact eigenstates of Hamiltonian (\ref{ham}), they still approximate well the exact edge states, especially for sufficiently large $u>1$ and $M =2N \gg 1$.
	In Fig.~\ref{Fig.1}(b) we show the wavefunctions of states $\ket{L,R}$ which are exponentially localized at the edges of the lattice with a localization length $\xi=2/\ln (u)$ (see below).
	
	Since the energies of the edge states $\ket{L,R}$ lie in the middle of the gap between the two bands, their interaction with the bulk eigenstates can be neglected. We can then write an effective two-state Hamiltonian for the edge states as 
	\begin{equation}
		\he=\begin{pmatrix}
			\braket{L|\h|L} & \braket{L|\h|R} \\
			\braket{R|\h|L} & \braket{R|\h|R} 
		\end{pmatrix} \equiv
		\begin{pmatrix}
			\mathcal{E}_L & C \\
			C^* & \mathcal{E}_R
		\end{pmatrix}.
		\label{Heff}
	\end{equation}
	 To evaluate the matrix elements of $\he$, we observe that, apart from that of the boundary terms $\ket{1}\bra{2}$ and $\ket{M-1}\bra{M}$, the contributions of all the other terms of $\h$ vanishes. We then obtain 
	\begin{equation}
		 \mathcal{E}_L = \mathcal{E}_R = 0, \quad C=v~\dfrac{1-u^{-2}}{1-u^{-M}}\left(-u\right)^{-(M/2-1)} .
		\label{heff}
	\end{equation}
	For $u > 1$ and $M \gg 1$, the effective coupling  between the two edge states can be approximated by  
	\begin{equation}
		|C|\simeq C_0~ e^{-M/\xi}, \label{coupl}
	\end{equation}
	where $C_0=\dfrac{w^2-v^2}{w}$ and $\xi=2/\ln (u)$, which implies  that the coupling falls off  exponentially with $M$ with a characteristic length $\xi$ which coincides with the localization length of the two edge states. This coupling is also responsible for the hybridization of the two edge states in finite systems. The hybridized states are the eigenstates of the effective Hamiltonian, $\he\ket{\pm}=E_{\pm}\ket{\pm}$, given by the symmetric and antisymmetric superpositions of the two edge states,
	\begin{gather}
		\ket{\pm}=\dfrac{1}{\sqrt{2}}\left( \ket{L}\pm\ket{R} \right) , \label{eq.LpmR}
	\end{gather}
	with the eigenenergies $E_{\pm} = \pm|C|$ that tend to zero in the limit of a long chain $M \gg \xi$.

	\section{$\mathcal{PT}$-symmetric SSH model}
	\label{Sec:PT}
	
	We now consider a $\mathcal{PT}$-symmetric extension of the SSH model governed by the non-Hermitian Hamiltonian 
	\begin{equation}
		\h_{\mathcal{PT}}=\h + i\gamma\sum_{m=1}^M (-1)^{(m-1)} \ket{m}\bra{m}, 
		\label{hampt}
	\end{equation}
	where $\gamma>0$ is the alternating gain and loss rate on the successive lattice sites, as shown schematically in Fig.~\ref{fig:spectr}(a). The gain-loss contrast $\gamma$ determines whether the $\mathcal{PT}$-symmetry is manifest in the real (Hermitial-like) spectrum, or is spontaneously broken \cite{PT0}. 
	In contrast to the Hermitian Hamiltonian $\h$, the Hamiltonian (\ref{hampt}) does not respect chiral symmetry but possesses pseudo-anti-Hermiticity, $\{\Sigma_z \mathcal{T},\h_{\mathcal{PT}}\}=0$, where $\mathcal{T}$ denotes complex conjugation that here coincides with the time-reversal operator. By construction,  $\h_{\mathcal{PT}}$ respects parity-time symmetry $[\hat{\mathcal{PT}},\h_{\mathcal{PT}}]=0$, where $\hat{\mathcal{PT}}$ is the $\mathcal{PT}$-symmetry operator, with $\mathcal{P}$ represented by the backward identity (exchange) matrix. The combined effect of these two symmetries leads to a spectrum symmetric with respect to both the real and the imaginary axis, $E\to E^*$ and $E\to -E^*$.

	\subsection{Bulk states}
	
	The bulk $k$-space Hamiltonian is now given by
	\begin{equation}
		\h_{\mathcal{PT}}(k)=\begin{pmatrix}
			i\gamma & v+w e^{-ik} \\
			v+w e^{ik} & -i\gamma
		\end{pmatrix} ,
	\end{equation}
	with the dispersion relation 
	\begin{equation}
		E(k)=\pm\sqrt{v^2+w^2+2vw\cos{k}-\gamma^2} 
	\end{equation}
	that leads to the $\mathcal{PT}$ classification for the bulk states of the system based on whether all, some, or none of the eigenvalues are real. The corresponding phases are
	\begin{align}
		\gamma<|v-w| &\qquad\text{unbroken}\nonumber \\
		|v-w|<\gamma<|v+w| &\qquad \text{partially broken} 	\label{eq:PT_clas} \\
		\gamma>|v+w| &\qquad \text{fully broken.}\nonumber 
	\end{align}
	It has been shown \cite{Esaki,Schomerus} that, as long as a bulk band gap exists, the topological properties of this system can be deduced again by a winding number of the same form as in the Hermitian case, Eq.~(\ref{eq:Windnum}).
	Hence, the system is topologically nontrivial for $w>v$ (such that $\mathcal{W}\ne0$) and $\gamma<|v-w|$ (open gap) or, equivalently, for $w>\gamma+v$.

	\subsection{Edge states}
	
\begin{figure}[tb]
		\centering
		\includegraphics[clip,width=\linewidth]{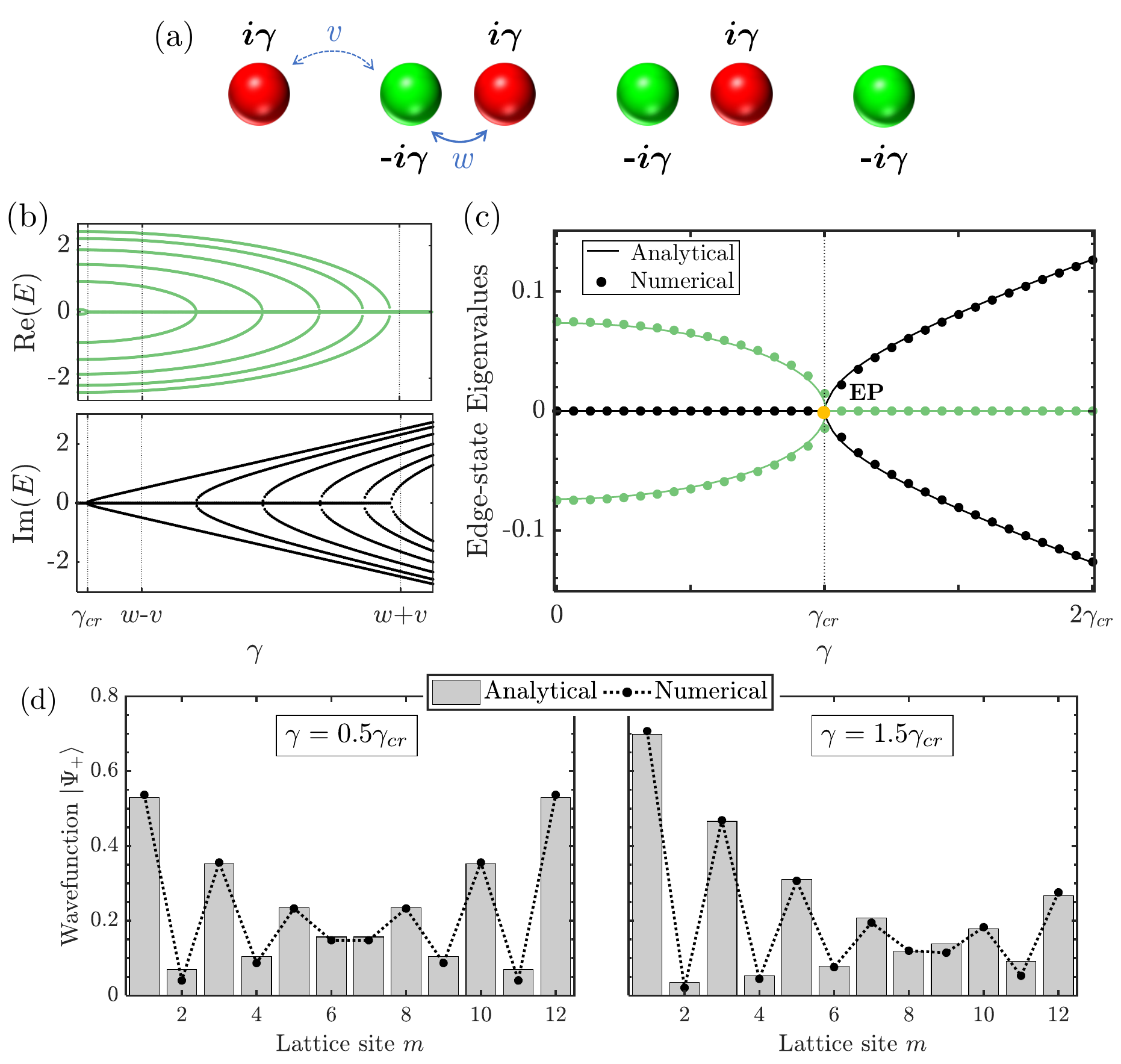}
		\caption{(a) Schematic of a $\mathcal{PT}$-symmetric SSH chain with the nearest-neighbor couplings $v < w$ and alternating gain and loss $\pm i \gamma$ [cf. Hamiltonian (\ref{hampt})].
		(b)~Real [top, dark-green (gray)] and imaginary (bottom, black) part of the energy spectrum of the chain of $M=12$ sites with $u=w/v = 1.5$ vs the gain-loss contrast $\gamma$. The energy $E$ is in units of $w$.
		(c)~Magnified view of (b) in the vicinity of $\gamma_{cr}\ll|v-w|$ where the edge-state eigenvalues  collapse to an exceptional point of Eq.~(\ref{gcr}).
		(d)~Spatial profile of state $\ket{\Psi_+}$  (absolute values of the amplitudes at lattice sites $m$) for $\gamma$ smaller (left) and larger (right) than  $\gamma_{cr}$; bars correspond to the analytical Eq.~(\ref{eq:PT_wavef}) with $\ket{L,R}$ of Eqs.~(\ref{ans}) and the black filled circles connected by dotted lines show the exact numerical results.}
		\label{fig:spectr}
	\end{figure}

	We now discuss the properties of the edge states in finite lattices with the topologically nontrivial configuration. 
	Note that the $\mathcal{PT}$ classification of Eqs.~(\ref{eq:PT_clas}) holds only in the thermodynamic limit, since for any finite system with nontrivial topology the edge-state eigenvalues collapse to an exceptional point at a small critical value of  $\gamma_{cr}\ll|v-w|$ as seen in Fig.~\ref{fig:spectr}. Observe also that, for short lattices ($M \sim 10$) with few eigenstates, the partially and fully $\mathcal{PT}$-broken phases of the bulk occur for the values of gain-loss contrast $|v - w|< \gamma < |v + w|$ \cite{Breakup}.
	
	Our main objective is to determine the value of $\gamma_{cr}$ at which the edge states attain an exceptional point and then acquire imaginary energy eingenvalues. To construct an effective two-state Hamiltonioan  $\he_{\mathcal{PT}}$, we use the same wavefunction ansatz of Eqs.~(\ref{ans}) for the edge states $\ket{L,R}$ and find that   
	\begin{equation}
	    \h_{\mathcal{PT}}\ket{L,R}=\h\ket{L,R}\pm i\gamma\ket{L,R}.
	\end{equation}
 	Since $\h\ket{L,R}=0$ for $M\to\infty$, the above relation reveals that the two states are also eigenstates of  $\h_{\mathcal{PT}}$ in the thermodynamic limit, but with the corresponding imaginary eigenvalues $\pm i \gamma$. This suggests that the ansatz~(\ref{ans}) is indeed suitable for the non-Hermitian case as well.
	 It is then easy to see that the diagonal elements (imaginary potential) in Eq.~(\ref{hampt}) will translate to diagonal elements $\mathcal{E}_{L,R}=\pm i \gamma$ of the effective Hamiltonian, but will not affect the off-diagonal elements of $\he_{\mathcal{PT}}$. Hence, the effective Hamiltonian is 
	\begin{equation}
		\he_{\mathcal{PT}}=\begin{pmatrix}
			i\gamma & ~C \\
			C & -i\gamma
		\end{pmatrix} ,
		\label{eq:Heff_PT}
	\end{equation}
	with $C$ given in Eq.~(\ref{heff}).
	The eigenvalues of $\he_{\mathcal{PT}}$ are  
	\begin{equation}
		E_{\pm}=\pm\sqrt{C^2-\gamma^2} , \label{eq:EPT}
	\end{equation}
	with the corresponding eigenstates 
	\begin{equation}
		\ket{\Psi_+}= \dfrac{1}{\sqrt{\mathcal{N}}}
		    \begin{pmatrix}
			\cos{\theta} \\
			\sin{\theta}
		\end{pmatrix}, \quad
		\ket{\Psi_-}= \dfrac{1}{\sqrt{\mathcal{N}}}
		\begin{pmatrix}
			\sin{\theta} \\
			-\cos{\theta}
		\end{pmatrix},  
		\label{eq:PT_wavef}
	\end{equation}
	where $\mathcal{N}$ a normalization constant and 
	$\theta\equiv \tfrac{1}{2}  \tan^{-1}{\left(\tfrac{C}{i\gamma}\right)}=-\tfrac{i}{4}\ln{\tfrac{\gamma+C}{\gamma-C}}$ is the (complex) mixing angle, which reduces to $\theta = \pi/4$ for $\gamma\to 0$ leading to the eigenstates $\ket{\pm}$ of the Hermitian case in Eq.~(\ref{eq.LpmR}).
	It then follows from Eq.~(\ref{eq:EPT}) that the critical value of the gain-loss contrast $\gamma_{cr}$ at which the eigenvalues of the edge-state are expected to turn from real to imaginary is 
	\begin{equation}
		\gamma_{cr}=|C|=v~\dfrac{1-u^{-2}}{1-u^{-M}}~u^{-(M/2-1)}.
		\label{gcr}
	\end{equation}
	
	In Fig.~\ref{fig:spectr}(c) we show the magnified spectrum of the system in the vicinity of $\gamma_{cr}$ obtained from the exact diagonalization of the full Hamiltonian (\ref{hampt}) and compare it with the analytical prediction of Eq.~(\ref{eq:EPT}). The corresponding wavefunctions for the eigenstate $\ket{\Psi_+}$ for two values of $\gamma$ below and above the $\mathcal{PT}$-symmetry breaking point are shown in Fig.~\ref{fig:spectr}(d). Notice that below the exceptional point, $\gamma<C$, the eigenstates of $\he_{\mathcal{PT}}$ have equal contributions from both edge states, $|\braket{L,R|\Psi_{\pm}}|=1/\sqrt{2}$, while above the exceptional point, $\gamma>C$, the states $\ket{\Psi_{\pm}}$ have increasingly larger contributions from $\ket{L,R}$, respectively, as expected. 
	
	\begin{figure}[t]
		\centering
		\includegraphics[clip,width=\linewidth]{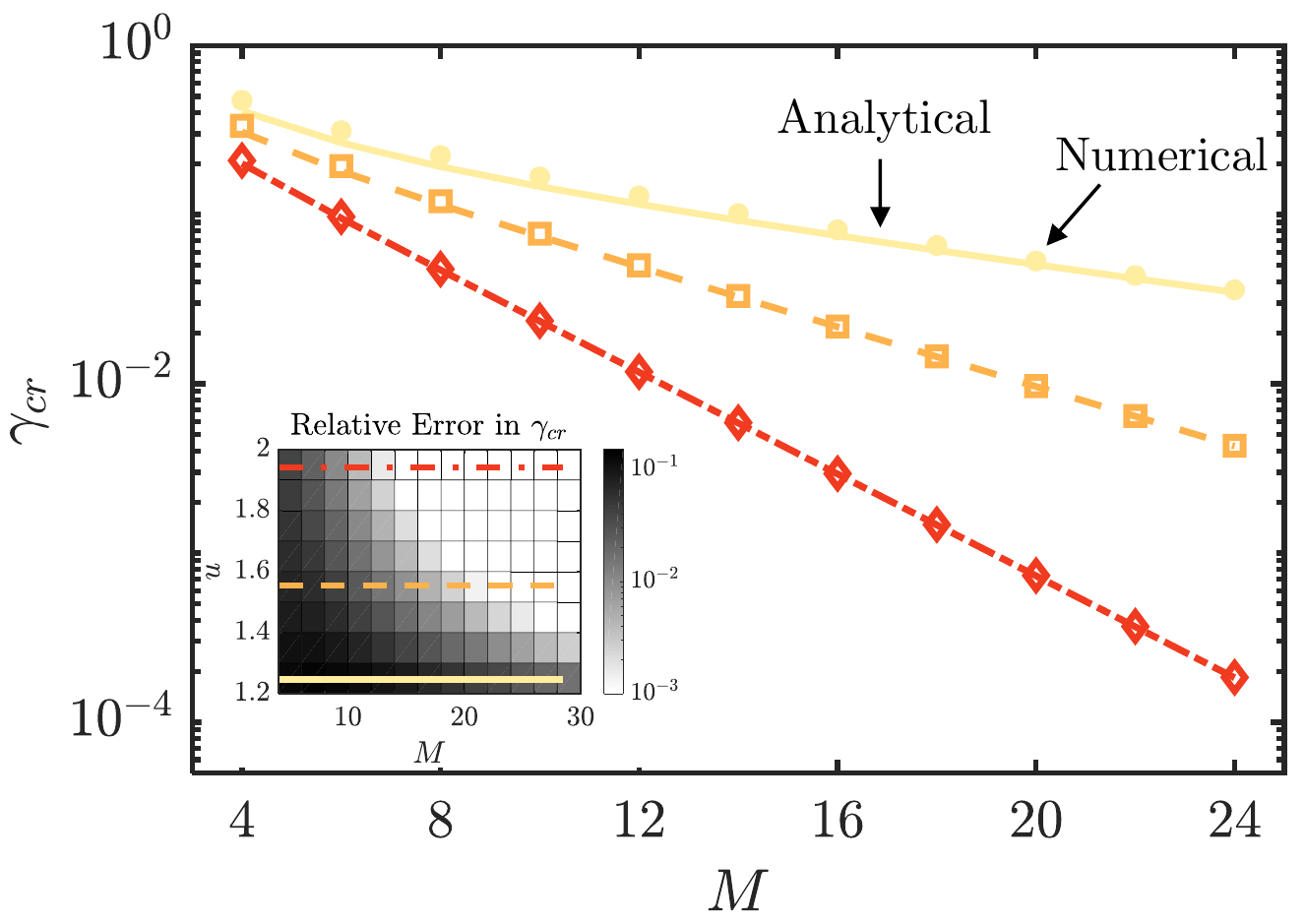}
		\caption{Critical values of the gain-loss contrast $\gamma_{cr}$ (in units of $w$) vs the lattice length $M$, for three different values of $u=w/v=1.2, 1.5, 2.0$. Main panel shows the comparison between the analytical expression (\ref{gcr}) (solid yellow, dashed orange, and dashed-dotted red lines) and the exact numerical values (circles, squares and diamonds). 
		Inset shows the relative error in $\gamma_{cr}$ of Eq.~(\ref{gcr}) vs $M$ and $u$, with the horizontal lines corresponding to the three values of $u$ used in the main panel. }
		\label{fig:gammacrit}
	\end{figure}
	
	In Fig.~\ref{fig:gammacrit} we compare  the analytical prediction of Eq.~(\ref{gcr}) with the values of $\gamma_{cr}$ obtained by exact diagonalization of the full Hamiltonian (\ref{hampt}) for different $M$ and $u=w/v$ and extraction of the exceptional points. We find nearly perfect agreement between the analytical and numerical results for sufficiently large $u \gtrsim 1.5$ and only small discrepancy for smaller values of $u$ and $M$. Note, finally, that, according to Eq.~(\ref{gcr}), $\gamma_{cr}$ decreases exponentially with increasing the system size $M$ and, hence, for sufficient large $M$ the exceptional point for the edge state eigenvalues occurs already in the close vicinity of $\gamma=0$.

	\section{$\mathcal{PT}$-symmetric SSH model with arbitrary gain and loss}
	\label{Sec:Random}
	
	With minor modifications, the effective description of the edge states presented above can be applied to a larger variety of systems with arbitrary gain and loss rates, with the only restrictions being that they must respect the global $\mathcal{PT}$ symmetry while retaining the bulk energy gap. 
	We therefore consider once again an SSH chain with alternating gain and loss rates $\gamma_{m}$ on the successive sites, as described by the Hamiltonian 
	\begin{equation}
		\tilde{\h}_{\mathcal{PT}}=\h+i\sum_{m=1}^M (-1)^{(m-1)} \gamma_{m}\ket{m}\bra{m},
		\label{eq:HPT_disorder}
	\end{equation}
	where $\h$ is the Hamiltonian of Eq.~(\ref{ham}), while the global $\mathcal{PT}$-symmetry requires that
	\begin{equation}
		\gamma_m=\gamma_{M-m+1}. \label{eq:globalPTsym}
	\end{equation}
	
    We again construct an effective two-state Hamiltonian for the edge states $\ket{L,R}$ of Eqs.~(\ref{ans}). As before, the off-diagonal elements of the effective Hamiltonian remain the same, $\braket{L|\tilde{\h}_{\mathcal{PT}}|R}=\braket{R|\tilde{\h}_{\mathcal{PT}}|L}^* = C$, while for the diagonal elements and we have 
	\begin{subequations}
	\begin{eqnarray}
	\mathcal{E}_L &\equiv & \braket{L|\tilde{\h}_{\mathcal{PT}}|L} = i \sum_{m=1}^{M} (-1)^{(m-1)}\gamma_{m} |c_m^L|^2 
	\nonumber \\ &=& i \bar{\gamma}, \\
	\mathcal{E}_R &\equiv & \braket{R|\tilde{\h}_{\mathcal{PT}}|R} = i \sum_{m=1}^{M} (-1)^{(m-1)} \gamma_{m} |c_m^R|^2 
	\nonumber \\ &=&
	-i \bar{\gamma} , \quad 
	\end{eqnarray}
	\label{eq:heff_diag}
	\end{subequations}
    where 
	\begin{equation}
		\bar{\gamma} \equiv \frac{\sum_{n=1}^N \gamma_{2n-1} \left(u^2\right)^{-(n-1)}}
		{\sum_{n=1}^N \left(u^2\right)^{-(n-1)}}
		\label{eq:gbcrit}
	\end{equation}
	is the effective gain-loss rate for states $\ket{L,R}$, respectively. Hence, the effective Hamiltonian is 
	\begin{equation}
		\tilde{\h}^{\mathrm{(eff)}}_{\mathcal{PT}}= \begin{pmatrix}
			i\bar{\gamma}& C\\
			C & -i\bar{\gamma}
		\end{pmatrix}, 
	\end{equation}
	with the eigenvalues (\ref{eq:EPT}) and eigenvectors (\ref{eq:PT_wavef}) with the replacement $\gamma \to \bar{\gamma}$.
	The critical value of $\bar{\gamma}$ at which the edge states attain an exceptional point are $\bar{\gamma}_{cr}=|C|$.
	
	\begin{figure*}[t]
		\centering
		\includegraphics[width=2.0\columnwidth]{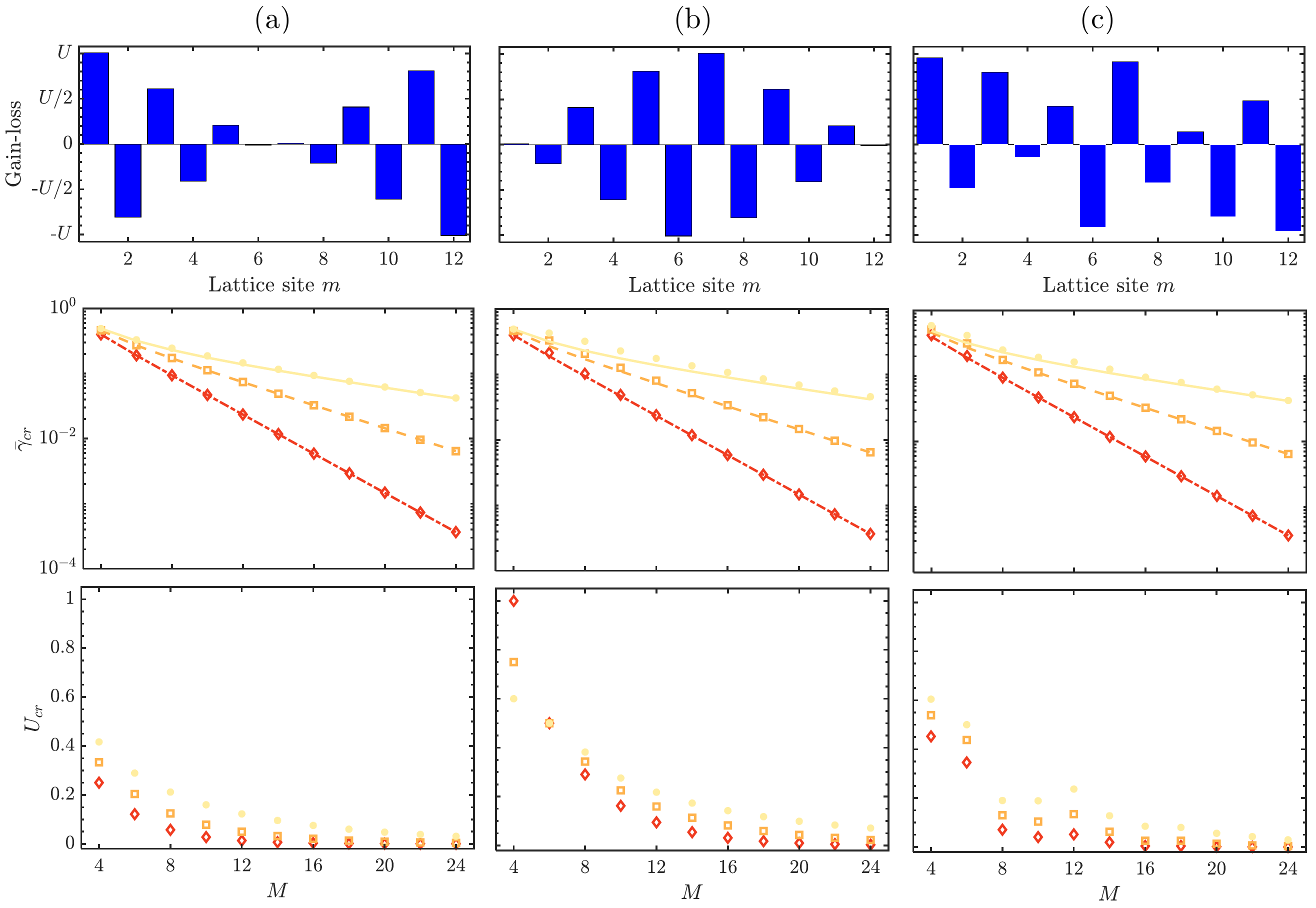}
		\caption{$\mathcal{PT}$-symmetry breaking in SSH chains with arbitrary gain and loss. 
		Top panels illustrate the three different complex potentials for the Hamiltonian (\ref{eq:HPT_disorder}) given by Eqs.~(\ref{eq:potentials}a,b,c), respectively. 
		Middle panels show the  critical values of $\bar{\gamma}_{cr}$ (in units of $w$) vs the chain length $M$ for $u=w/v=1.2,1.5,2.0$ (same line and symbol styles and color code as in Fig.~\ref{fig:gammacrit}) obtained numerically (symbols) and analytically (lines). 
		Bottom panels shown the critical values of the potential amplitudes $U$ (in units of $w$) in Eqs.~(\ref{eq:potentials}) that correspond to $\bar{\gamma}_{cr}$ in the middle-panel plots.}
		\label{fig:4}
	\end{figure*}
	
	To verify our analytical predictions for the general model in Eq.~(\ref{eq:HPT_disorder}), we consider three different spatial distributions of the complex potential $\gamma_m \in \, [0, U]$ for $m=1,2,\ldots , M/2$, 
	\begin{subequations}
	\begin{eqnarray}
	\gamma_{m} &=& U \, \frac{M/2 - m}{M/2-1} , \\
	\gamma_{m} &=& U \, \frac{m - 1}{M/2-1} \\
	\gamma_{m} &=& U \, r_m ,
	\end{eqnarray}
	\label{eq:potentials}
	\end{subequations}
    which satisfy Eq.~(\ref{eq:globalPTsym}) for $m=M/2 + 1, M/2 + 2,\ldots, M$. 
	Here case (a) corresponds to $\gamma_m$ linearly decreasing from the edges to the center of the chain; 
	case (b) corresponds to $\gamma_m$ linearly increasing from edge to center; and case
	(c) corresponds to random $\gamma_m$ with uniformly distributed $r_m \in [0,1]$; as we illustrate in the top panels Fig.~\ref{fig:4}. 
	In the middle panels of Fig.~\ref{fig:4} we plot the critical values of $\bar{\gamma}_{cr}$ at which the edge-state eigenvalues collapse to an exceptional point, as predicted analytically, $\bar{\gamma}_{cr}=|C|$, and obtained numerically via exact diagonalization of the full Hamiltonian (\ref{eq:HPT_disorder}) for different $M$ and $u=w/v$. In all cases, we find good agreement between the analytical and numerical calculations. The largest deviation between the analytical and numerical values of $\bar{\gamma}_{cr}$ are obtained for small $u\lesssim 1.5$ and $M \lesssim 10$, especially for the case (b) with the maximum of the potential in the middle of the chain, $\gamma_{M/2} = U$. 
	In the bottom panels of Fig.~\ref{fig:4} we show the values of $U=U_{cr}$ corresponding to the exceptional points ($\bar{\gamma} = \bar{\gamma}_{cr}$) obtained from the numerical simulations. $U_{cr}$ is largest for case (b) since the wavefunctions of the edge states that decay away from the boundaries are less affected by the stronger imaginary potential in the middle of the chain. Now the potential in the vicinity of $m=M/2$ can take large values, $\gamma_{m} \sim |w-v|$, comparable to the bulk gap. But the strong potential in the middle of the chain significantly affects the spectrum of the bulk which in turn can perturb the zero-energy edge states. This explains the largest deviation between the analytically predicted and numerically calculated values of $\bar{\gamma}_{cr}$ for case (b) with small $u$ and $M$ for which the edge state wavefunctions have relativelly large amplitudes at the middle of the chain.       
	
	\section{Conclusions}
	\label{Sec:Concl}
	
	To summarize, we have studied the single-particle, non-Hermitian, parity-time ($\mathcal{PT}$) symmetric Su-Schrieffer-Heege (SSH) model in the topologically non-trivial configuration and derived an effective analytical model for edge states in finite lattices. Our effective model accounts for the evanescent coupling between the edge states, accurately describes their properties, and gives physically transparent interpretation for the  $\mathcal{PT}$-symmetry breaking for the edge states, which we verified by exact numerical calculations. 
	
	The studied system has experimental relevance as both the ordered and disordered non-Hermitian Hamiltonians (\ref{hampt}) and (\ref{eq:HPT_disorder}) can be physically implemented by the addition of alternating optical gain and loss in a lattice of coupled optical elements, such as waveguides or cavities, as has been realized in a number of recent experiments \cite{NhTopo1,Makris1}. 
	
	\acknowledgments
	A.F.T., N.E.P. and D.P. were supported by the EU QuantERA Project PACE-IN (GSRT Grant No. T11EPA4-00015).

\end{document}